\documentclass[fleqn,usenatbib]{mnras}

\usepackage{newtxtext,newtxmath}

\usepackage[T1]{fontenc}

\DeclareRobustCommand{\VAN}[3]{#2}
\let\VANthebibliography\thebibliography
\def\thebibliography{\DeclareRobustCommand{\VAN}[3]{##3}\VANthebibliography}

\usepackage{graphicx}	
\usepackage{amsmath}	

\title[SMILE: Search for MIlli-LEnses]{SMILE: Search for MIlli-LEnses}

\author[C.~Casadio et al.]{
C.~Casadio,$^{1,2,3}$\thanks{E-mail: ccasadio@ia.forth.gr (CC)}
D.~Blinov,$^{1,2,4}$
A.~C.~S.~Readhead,$^{5}$
I.~W.~A.~Browne,$^{6}$
P.~N.~Wilkinson,$^{6}$
T.~Hovatta,$^{7,8}$ \newauthor
N.~Mandarakas,$^{1,2}$
V.~Pavlidou,$^{1,2}$
K.~Tassis,$^{1,2}$
H.~K.~Vedantham,$^{9,10}$
J.~A.~Zensus,$^{3}$
V.~Diamantopoulos,$^{2}$ \newauthor
K.~E.~Dolapsaki,$^{2}$
K.~Gkimisi,$^{2}$
G.~Kalaitzidakis,$^{2}$
M.~Mastorakis,$^{2}$
K.~Nikolaou,$^{2}$
E.~Ntormousi,$^{11,1,2}$ \newauthor
V.~Pelgrims$^{1,2}$
and K.~Psarras$^{2}$
\\
$^{1}$Institute of Astrophysics, Foundation for Research and Technology - Hellas, Voutes, 7110 Heraklion, Greece\\
$^{2}$Department of Physics, University of Crete, 71003, Heraklion, Greece\\
$^{3}$Max-Planck-Institut f\"{u}r Radioastronomie, Auf dem H\"{u}gel 69, 53121 Bonn, Germany\\
$^{4}$St. Petersburg State University, Universitetsky pr. 28, Petrodvoretz, 198504 St. Petersburg, Russia\\
$^{5}$Cahill Center for Astronomy and Astrophysics, California Institute of Technology, 1200 E California Blvd, MC 249-17, Pasadena CA, 91125, USA\\
$^{6}$University of Manchester, Jodrell Bank Observatory, Nr. Macclesfield, Cheshire SK11 9DL\\
$^{7}$Finnish Centre for Astronomy with ESO, FINCA, University of Turku, Finland\\
$^{8}$Aalto University Mets\"ahovi Radio Observatory, Mets\"ahovintie 114, FI-02540 Kylm\"al\"a, Finland\\
$^{9}$Kapteyn Astronomical Institute, University of Groningen, The Netherlands\\
$^{10}$Leiden Observatory, Leiden University, PO Box 9513, 2300 RA Leiden, The Netherlands\\
$^{11}$Scuola Normale Superiore, Piazza dei Cavalieri 7, I-56126 Pisa, Italy
}

\date{Accepted XXX. Received YYY; in original form ZZZ}

\pubyear{2021}

\begin{document}
\label{firstpage}
\pagerange{\pageref{firstpage}--\pageref{lastpage}}
\maketitle

\begin{abstract}
Dark Matter (DM) halos with masses below $\sim 10^{8} M_{\sun}$, which would help to discriminate between DM models, may be detected through their gravitational effect on distant sources. The same applies to primordial black holes, considered as an alternative scenario to DM particle models. However, there is still no evidence for the existence of such objects. With the aim of finding compact objects in the mass range $\sim$ 10$^{6}$ -- 10$^{9} M_{\sun}$, we search for strong gravitational lenses on milli (mas)-arcseconds scales (< 150 mas). For our search, we used the Astrogeo VLBI FITS image database -- the largest publicly available database, containing multi-frequency VLBI data of 13828 individual sources. We used the citizen science approach to visually inspect all sources in all available frequencies in search for images with multiple compact components on mas-scales. At the final stage, sources were excluded based on the surface brightness preservation criterion. We obtained a sample of 40 sources that passed all steps and therefore are judged to be milli-arcsecond lens candidates. These sources are currently followed-up with on-going European VLBI Network (EVN) observations at 5 and 22 GHz. Based on spectral index measurements, we suggest that two of our candidates have a higher probability to be associated with gravitational lenses.
\end{abstract}

\begin{keywords}
gravitational lensing: strong -- techniques: interferometric -- quasars: general -- dark matter
\end{keywords}



\section{Introduction}
The distribution of mass aggregates below $\sim$10$^{9} M_{\sun}$ is of crucial importance in structure formation and possibly also in identifying the nature of Dark Matter (DM).  The standard cold dark matter model (CDM) predicts the formation of a larger number of DM halos on sub-galactic scales (subhalos) than  the warm dark matter model (WDM). Given that DM halos with masses below $\sim 3 \times 10^{8} M_{\sun}$ should not form galaxies \citep{Benitez2020}, the only way to detect them in distant galaxies, is through gravitational lensing effects on distant background sources \citep[e.g.,][]{Vegetti2012}. 

Gravitational lensing occurs when light rays from a distant source are bent by the gravitational potential of a foreground mass distribution ({\it lens}). If the surface density of the lens is high enough \citep{Treu2010,Zackrisson2010}, and if a compact background source and the lens are well-aligned with  the observer, multiple images of the background source will be formed. This is the case of {\it strong lensing}.  Based on the angular scale at which multiple images form in the lens plane, strong lensing can be further divided into three regimes: {\it macro} ($\sim$ arcseconds), {\it milli} (milli-arcseconds) and {\it micro} (micro-arcseconds) lensing.   

The cosmic lens all sky survey (CLASS) \citep{Myers2003,Browne2003} and Jodrell VLA astrometric survey (JVAS) \citep{King1999} used the Very Large Array (VLA) to identify 20 new gravitational macro-lenses due to galactic mass lenses $\sim$10$^{11} M_{\sun}$, and found a lensing rate of 1 macro-lens per 690$\pm$190 radio sources observed in a complete flux density limited sample  \citep{Browne2003}.

Milli-arcsecond gravitational lenses or {\it milli-lenses} can probe different astrophysical systems. Lensed images separated by 5 -- 100 mas are expected when the lens is a compact object (CO) with mass between 10$^{6}$ -- 10$^{9} M_{\sun}$, i.e., a \textit{supermassive CO}. Hence, milli-lenses can reveal dormant supermassive black holes located at the centers of galaxies or free-floating in the intergalactic space. In this case the black hole splits the image formed close to the center of mass \citep{Mao2001, Winn2004, Mueller2020} into multiple images on mas-scales, which are accessible to very long baseline interferometry (VLBI) observations. 

\cite{Wilkinson2001} used VLBI multi-frequency observations of 300 sources to search for milli-lenses with multiple-image separations in the angular range 1.5-50 milliarcseconds. Following \cite{Press1973}, their null result allowed them to constrain the CO mass density ($\Omega_{CO}$) for CO in the mass range $10^{6} - 10^{8} M_{\sun}$, to $\Omega_{CO}$ < 0.01 (95$\%$ confidence).

The Very Long Baseline Array (VLBA) has been used by \citet{Spingola2019} at 1.4 GHz in a search for milli-lenses with image separations $>$ 100 mas, corresponding to lensing mass scales $> 3 \times 10^9 M_{\sun}$. In terms of mass scale this complements the study we report here.  Following \cite{Wilkinson2001}, we selected a higher frequency than 1.4 GHz in order to focus on lensing of the compact cores of blazars and to reduce the confusing effects of the radio jets, which have steeper spectra than the cores.
The cores we are targeting for multiple images due to lensing are much more prominent relative to the jets at 8 GHz than at 1.4 GHz.

Lately there has been a renewed interest in the primordial black hole (PBH) scenario, partially due to the merging of a $\sim$85 $M_{\sun}$ and a $\sim$66 $M_{\sun}$ black hole, recently detected by the LIGO/Virgo Collaboration \citep{DeLuca2020arXiv200901728D}. If PBHs exist they might be the primary constituent of DM \citep[e.g.,][]{Clesse2018}. Moreover, given the wide range of possible masses of these objects (10$^{-18} M_{\sun}$ -- 10$^{9} M_{\sun}$, \citealp{Sasaki_2018}) PBHs could explain the high BH masses in the above LIGO/Virgo event. 

In this paper, we present a pilot search for milli-lenses with angular separation < 150 mas, that we carried out using the Astrogeo VLBI FITS image database\footnote{\url{http://astrogeo.org/vlbi\_images}}. We used VLBI data of all 13828 sources, available when this study was performed. A similar study was presented by \cite{Burke-Spolaor2011}. There, the search for binary supermassive black holes using spectral index images of 3114 sources, produced only one candidate. 
The bulk of sources in the Astrogeo VLBI FITS image database come from dedicated astrometry programs devoted to the completion of a radio fundamental catalog \citep{Petrov2021}. For this reason Astrogeo contains mostly compact radio sources observed in two frequency bands, and one of them is usually the X-band (8 GHz). This relatively high frequency is ideal for gravitational lens searches since at this frequency most of these sources are dominated by an unresolved flat-spectrum core, making identification of multiple lensed images of the core straightforward.

We identify 40 milli-lens candidates, which are promising targets for follow-up observations. We additionally provide an example of such follow-up, using newly obtained EVN data, for two of our candidates. We encourage further observations of this kind for the rest of our 40 candidates.

\section{Search for lenses}\label{search}

For this study, we downloaded all the UV-FITS files, with associated clean components, available at the Astrogeo VLBI FITS image database at the time this search was performed, for a total of 13828 individual sources. Starting from self-calibrated visibilities and using clean components, we produced new images with circular restoring beams of diameter d=$\sqrt{a_{maj}\times a_{min}}$, where $a_{maj}$ and $a_{min}$ are the major and minor axis of the restoring beams in the original image FITS files. In case of sources with multi-epoch observations at any particular observing band, we convolved visibilities with a median circular beam and we median-stacked the multi-epoch images together, using intensity peaks to align them. Final images have a dimension of 1024 $\times$ 1024 pixels and 0.3 mas/pixel, resulting in a field of view of $\sim$307 $\times$ 307 mas.

In order to visually inspect all resulting images of all sources, we then created a web-page similar, in concept, to interfaces of citizen-science projects \citep[e.g.,][]{Banfield2015}. At every reload the web-page showed the final images in all available bands for a single source. Images were displayed using JS9\footnote{\url{https://js9.si.edu/}} windows which allowed us to change flux density levels, dynamic range, zoom, etc. Hyperlinks to original single epoch FITS and poscript image files were also included in the web-page. At the bottom of the web-page, two buttons, "Lens" and "No Lens" allowed the user to judge the source as a possible gravitational lensing candidate or not, on the basis of the visual inspection of displayed images. As a control we also inserted, within the sample, 200 mock lens images, i.e. images displaying two separated compact components, that were generated in the following way. First we visually selected 200 random sources from the sample where only a central compact source and no secondary components within the 307$\times$307 mas field of view were present at any available band. Then we fitted the source core with a circular Gaussian function and randomly scaled its parameters to create the mock lens. The height and the standard deviation of the Gaussian were multiplied by random factors drawn from uniform distributions in the range [0.3, 1] and [0.8, 1.2] respectively. After this, we added the corresponding mock Gaussian function to a random location in the image. If multiple bands were present the same scaling factor was used for all of them. We repeated this procedure for all the 200 sources. In this way the mock lens images were reproduced as double compact components with similar spectral index and surface brightness. We mixed these mock lensed objects in the database so that they would be offered for classification by users in a random order among real sources. The mock lenses were injected into the database in order to evaluate attentiveness of each user and roughly estimate completeness of our search. Users were instructed to select any sources with multiple compact components regardless their flux ratio, surface brightness or separation. For this reason we did not aim to create a set of mock lenses, whose parameters would follow physically realistic distributions. Moreover, given that the lens mass distribution is unknown and no milli-lenses have been detected so far, it may not be possible to create such a set.

The first step of the sample selection consisted in the visual inspection of the 13828 sources through the web-page interface. To complete this task, 5 PhD scientists and 9 undergraduate Physics students from the University of Crete were involved. They were asked to mark as "Lens" sources showing multiple compact components in at least one of the available observing band, and as "No Lens" the rest of sources. Considering the lack of expertise in VLBI images of AGN for most of the participants in the search, people were asked to mark also as "Lens", sources with some kind of extension (even core-jet structure) in case of doubt on the nature of the source itself. In this way we sought to prevent loss of potential candidates right at the start of the search. Each source in the database was inspected by one user and the corresponding choice was recorded. After this first step, 950 sources (excluding the mock lenses) had been cataloged as potential lenses. Among the 200 mock lenses, 8 were missed. Two specific persons together were responsible for missing six out of these 8 sources. Given the higher than average rate of failure, their judgments were considered unreliable. Therefore, two of the authors who are experienced VLBI researchers re-inspected all sources classified by the original two people. They recovered the mock lenses and also selected 4 more lens candidates in addition to the 950 previously obtained. The final result is that the number of missed mock lenses was two and hence we estimate the final loss rate of real lens candidates as $\sim 1 \%$ (2/200).

After this first stage, a detailed visual inspection of the 954 sources was performed by CC and DB. Sources were rejected if they showed clear surface brightness or spectral index discrepancies between components, clear processing artifacts and/or clear core-jet structures without evidence of any other unresolved bright component aside from the core. After this second stage we were left with 128 candidates. Three more authors (AR, IB, and PW), who led the CLASS project and were therefore familiar with radio images of lenses, then examined these 128 sources, and rejected 69 more objects.
Our remaining sample, after these multiple stages of visual inspections, consisted of 59 lens
candidates.

Since a significant number of images come from astrometry and geodesy programs, with poor $uv$-coverage and short integration
times, they may have artifacts and lack extended weak emission. Therefore, we decided to use a conservative approach, and finally reject sources based on the surface brightness preservation criterion, i.e., components fainter than the core should appear more compact.
Since gravitational lensing preserves the surface brightness of the background source, we expect the lensed images to have all the same intrinsic brightness. In \cite{Browne2003} a apparent surface brightness ratio of $\sim$4 between the putative lensed images has been used as a conservative upper limit to select lens candidates within the CLASS program. Here instead, given the wide range of image quality, we used a less stringent apparent surface brightness ratio $<$ 7 to select our final list of gravitational lens candidates, 40 in total. 

\section{Best candidates and optical counterparts}\label{results}

\begin{figure}
   \center
    {\includegraphics[width=0.49\hsize]{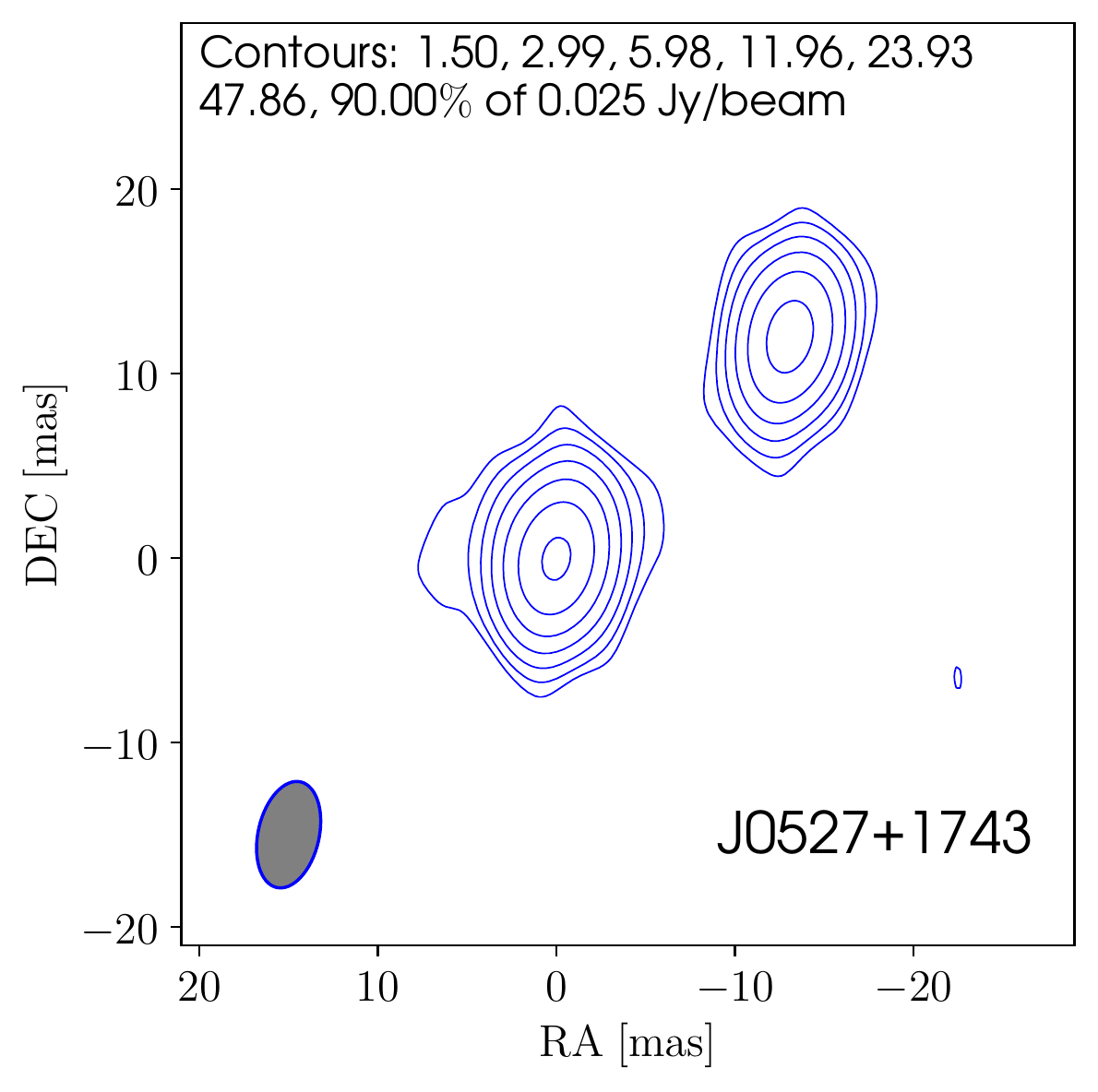}
     \includegraphics[width=0.49\hsize]{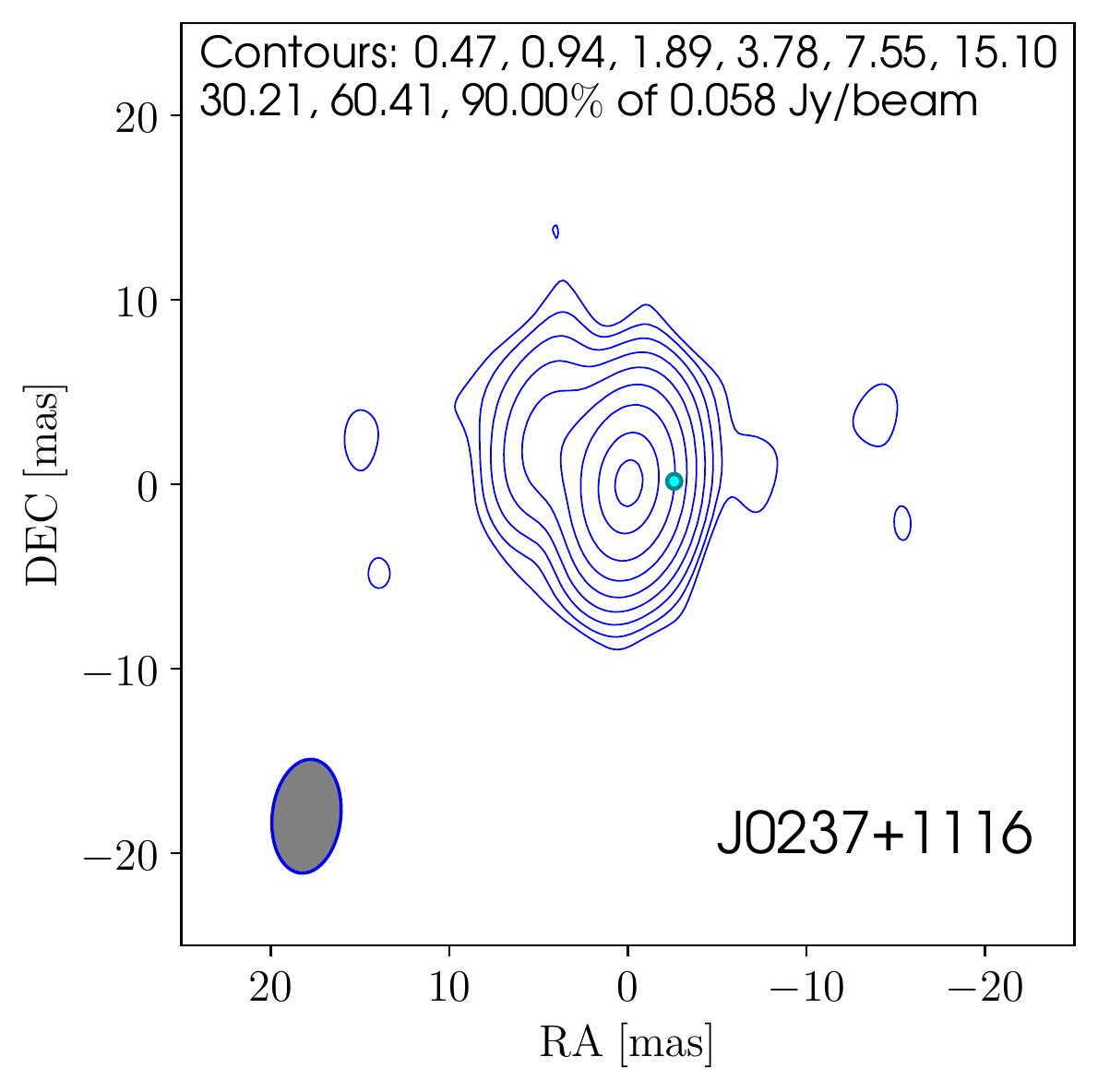}}
      \caption{New EVN 5 GHz images of two milli-lens candidates, shown as example. The first contour is at 3 $\times$ rms of the map noise. Grey ellipses are the respective restoring beams. The cyan dot indicates the optical photocenter position \citep{Kovalev2020}, with errorbars taken from Gaia EDR3 that are smaller than the symbol size. In the new EVN image, J0527+1743 preserves the multi compact components structure, while J0237+1116 shows a clear core-jet structure, which differs from the image selected within our search for milli-lenses.}
         \label{Gaia-VLBI}
   \end{figure}

The final 40 gravitational lens candidates which passed the multiple-stage selection are listed in Table~\ref{table:1}. This means that all 40 sources show multiple compact components at least in one of the available frequency and the apparent surface brightness ratio between components is $<$ 7. In Fig.~\ref{Gaia-VLBI} we show two of the milli-lens candidates as they appear in our new EVN 5GHz observations, as an example. The completion of the remaining sources' observations and data reduction is on-going. 

If the multiple components are the putative lensed images of the same source, we expect them to have the same spectra and flux density ratio between epochs. Given the small time delays between images in millilenses (e.g., characteristically around 20 seconds for a CO of 10$^{7} M_{\sun}$, \citep{Nemiroff2001}), intrinsic variability on such short timescales is likely to have negligible effect on flux density ratios and spectral indices.   
For all but five sources (J0213+8717, J1143+1834, J1632+3547, J1653+3503, and J1805-0438) we have a contemporary observation at two different frequencies (either 8 GHz and 4.3 GHz or 8 GHz and 2.3 GHz) that we use to compute the spectral indices of the multiple components. We used {\tt Difmap} \citep{shepherd1997difmap} to fit the multiple components observed in each source with circular Gaussian brightness distributions and we computed the components spectral indices $\alpha$ (S$_\nu\propto\nu^\alpha$) at the available frequencies.  As pointed out by \cite{Spingola2019}, spectral indices may be biased toward high negative values because of the combined effect of higher angular resolution and poorer sensitivity at high frequencies. Moreover, we cannot discard the possibility that light rays, which go through different paths to produce multiple images, cross different plasma with different absorption coefficients. This would also lead to multiple images with different spectral indices.
Therefore, we use the spectral indices obtained only to suggest, among the 40 best candidates, sources that have the highest probability to be associated with gravitational lenses rather than reject candidates. These must have multiple components with similar flat ($\alpha < 0.5$) spectral indices, in order to be lensed images of a core-jet system in the background. Among the 40 candidates, the two sources which best satisfy these criteria are J0527+1743, and J2312+0919. A third promising source is J1143+1834, for which no spectral index information is available.  

We also investigated the optical counterparts of the 40 lens candidates, using the publicly available optical catalogs in \textit{Aladin Sky Atlas}. Only 15 sources have associated optical counterparts and among them, only 4 sources (J0024-4202, J0210-2213, J2214-2521, J2312+0919) have known redshifts.

Another important piece of information comes from the comparison between the optical and VLBI position. VLBI-\textit{Gaia} offsets are a widely discussed topic \citep[e.g.,][and references therein]{Kovalev2017,Xu2021} and it has been shown that such offsets can be used as a tool to distinguish between the accretion disk and jet emission regions \citep{Plavin2019}. Six sources from our sample have coordinates within 0.5 arcseconds from optical sources in the \textit{Gaia Early Data Release (EDR) 3} \citep{gaiacollaboration2020gaia}. However, only for two of them accurate VLBI-Gaia offsets are available. In J0237+1116, the \textit{Gaia} position is shifted in the opposite direction of the extended emission in C-band which connects the two bright components at X-band, and is located 2.6 mas west the brightest component \cite{Kovalev2020}. This may suggest that what we see in our VLBI image is a core-jet structure while the optical emission comes mostly from the accretion disk following \cite{Plavin2019}. This is supported by our newly obtained EVN 5 GHz image (see Fig.~\ref{Gaia-VLBI}), where the source shows a clear core-jet structure, which allows us to discard it as a lens. In J2214-2521, for which the radio position is available in ICRF3 \citep{Charlot2020}, the VLBI-Gaia offset is $\sim$155 mas. The separation between radio components in this source is an order of magnitude smaller. The optical image from PanSTARRS which reveals the presence of an extended galaxy at redshift z=0.0868, suggests that in this case some bright optical emitting region within the galaxy may be responsible for the offset of the optical position or, an alternative scenario, the radio source is a background object, while \textit{Gaia} measures the optical photocenter of the nearby foreground galaxy. We will further investigate VLBI-Gaia offsets of our lens candidates in future publication after a careful analysis of EVN phase-referencing observations.

\begin{table}
\caption{Best lens candidates: (1) - J2000 name; (2) and (3) - Right ascension and Declination (J2000) according to NASA NED.}
\label{table:1}
\centering
\begin{tabular}{l c c}
\hline\hline
ID & RA    & DEC \\
   & h:m:s & d:m:s \\
\hline
J0010$-$0740 &  00:10:50.60 & $-$07:40:12.10  \\ 
J0011+3443   &  00:11:17.01 &  +34:43:33.64   \\ 
J0024$-$4202 &  00:24:42.99 & $-$42:02:03.95  \\
J0044+2858   &  00:44:21.55 & +28:58:33.84     \\
J0052+1633   &  00:52:36.17 & +16:33:00.45    \\ 
J0118+3810   &  01:18:10.14 & +38:10:55.10   \\
J0132+5211   &  01:32:18.92 & +52:11:30.70    \\
J0139+0824   &  01:39:57.16 & +08:24:26.80   \\
J0203+3041   &  02:03:45.36 & +30:41:29.11    \\
J0210$-$2213 &  02:10:10.06 & $-$22:13:36.90  \\
J0213+8717   &  02:13:57.85 & +87:17:28.8    \\
J0222+0952   &  02:22:15.61 & +09:52:37.80 \\
J0232$-$3422 &  02:32:30.03 & $-$34:22:03.10  \\
J0237+1116   &  02:37:13.59 & +11:16:15.48  \\
J0502+1626   &  05:02:47.39 & +16:26:39.32  \\
J0527+1743   &  05:27:23.21 & +17:43:25.10  \\
J0616$-$1957 &  06:16:01.57 & $-$19:57:16.20 \\
J0732+6023   &  07:32:50.97 & +60:23:40.06   \\
J0923$-$3435 &  09:23:53.88 & $-$34:35:26.10 \\
J1132+5100   &  11:32:50.39 & +51:00:19.92 \\
J1143+1834   &  11:43:26.07 & +18:34:38.36 \\
J1218$-$2159 &  12:18:58.82 & $-$21:59:45.4 \\
J1306+0341   &  13:06:16.00 & +03:41:40.80  \\
J1340$-$0335 &  13:40:13.30 & $-$03:35:20.80 \\
J1344$-$1739 &  13:44:03.42 & $-$17:39:05.50 \\
J1632+3547   &  16:32:31.25 & +35:47:37.74  \\
J1653+3503   &  16:53:53.16 & +35:03:27.03  \\
J1721+5207   &  17:21:36.26 & +52:07:10.40 \\
J1805$-$0438 &  18:05:31.12 & $-$04:38:09.69 \\
J2010+1513   &  20:10:08.20 & +15:13:58.84  \\
J2044+6649   &  20:44:49.19 & +66:49:02.30 \\
J2114+4036   &  21:14:10.01 & +40:36:42.19 \\
J2209+6442   &  22:09:30.49 & +64:42:20.70  \\
J2214$-$2521 &  22:14:46.39 & $-$25:21:16.00 \\
J2225+0841   &  22:25:43.48 & +08:41:57.20  \\
J2259+4037   &  22:59:04.04 & +40:37:47.10  \\
J2312+0919   &  23:12:28.07 & +09:19:26.70 \\
J2324$-$0058 &  23:24:04.62 & $-$00:58:54.20 \\
J2337$-$0622 &  23:37:29.19 & $-$06:22:13.20 \\
J2347$-$1856 &  23:47:08.63 & $-$18:56:18.86 \\
\hline
\end{tabular}
\end{table}

We report below a few extra notes and considerations on some of the sources. 

\textit{J0024-4202}: Previously classified as a Gigahertz Peaked Spectrum (GPS) radio galaxy \citep[e.g.,][]{Labiano2007}, the optical counterpart is located at redshift z=0.937.

\textit{J0210-2213}: Classified as a GPS source \citep{Snellen2002}; the optical counterpart is located at redshift z=1.49. 

\textit{J1132+5100}: Located behind the galaxy cluster Abell~1314 \citep{Vallee1989}. The different spectral indices of components A and B suggest that they are not images of each other and should be discarded. However, the peculiar location of this source, behind a galaxy cluster, is favorable for the search of gravitational lens systems. 

\textit{J1143+1834}: This source has been previously reported as Compact Symmetric Object (CSO) \citep{Sokolovsky2011} and Supermassive Binary Black Hole (SBBH) candidates \citep{Tremblay2016}.

\textit{J1632+3547}: \cite{Tremblay2016} discarded this source as a CSO candidate, but they still propose it as SBBH candidate.

\textit{J1653+3503}: This source has been discarded as both CSO and SBBH candidate \citep{Tremblay2016}.

\textit{J2312+0919}: Associated with a Fanaroff-Riley II (FRII) - high excitation radio galaxy, 3C~456, \citep[e.g.,][]{Macconi2020}; the optical counterpart is located at redshift z=0.233.

\textit{J2347-1856}: Previously classified as a CSO   \citep{Taylor2003, Sokolovsky2011}.

\section{Candidate confirmation}

We performed a pilot search for milli-arcsecond gravitational lenses to probe the existence of COs in the mass range of 10$^{6}$ -- 10$^{9} M_{\sun}$. The possible locations and origins of such objects, are different. For example, a supermassive CO could be a quiescent supermassive black hole hosted within a galaxy, or one of the CDM subhalos expected to surround the host galactic CDM halo \citep{Kravtsov2010}. In both cases, we should expect to see optical  emission associated to the host galaxy in the lens plane.
In the case of a free-floating CO (either a supermassive black hole or a DM halo), it is possible that no optical emission is associated with the lens. Hence, the optical emission, if present, should be associated with the lensed source and, therefore, should be lensed like the radio emission. 
This means that, if the radio and optical are co-spatial in the lensed source, the \textit{Gaia} position is expected to be in between the putative lensed images, given the inability of \textit{Gaia} observatory to resolve components with angular separation lower than $\sim$ 100 mas \citep{Lemon2017}. The situation is different if the optical and radio emissions are not co-spatial \citep[e.g.,][]{Reines2020}. In this case, we may still expect the optical emission being lensed, but the \textit{Gaia} position is not expected necessarily to fall in between the putative lensed images observed at VLBI scales.

The optical counterparts of these sources will be one of the aspects to investigate in further analysis, but we mostly expect multi-frequency and/or multi-epoch VLBI observations in future to be fundamental for finally confirming or rejecting the lens nature of these candidates. Since the short time delays in milli-lenses do not create delay-dependent changes in flux ratios, multi-epoch observations will help in discarding many candidates based on flux density ratio measurements between the putative lens images. Ultimately, the sources that will preserve flux density ratios despite flux density variability over epochs, and have multiple compact components with similar spectral indexes, will be tested using lens modelling to verify if the configuration and relative flux densities are compatible with a gravitational lens system.

\section{Discussion and further actions}\label{conclusions}

The sample of the 40 most probable milli-lens candidates is being followed-up with EVN observations at 5 and 22 GHz (Project ID: EC071; PI: Casadio). Higher sensitivity 5 GHz images should help to discard core-jet structures if present, and with the addition of the 22 GHz we hope to better resolve the structure of compact components and to add more spectral information.
We encourage the community to join with more follow-up observations. 

Any sources that do not pass the follow-up tests and are rejected as milli-lens candidates can still be investigated as candidates for a  supermassive binary black hole system or a CSO.
CSO and GPS radio galaxies, based on morphological and spectral classification respectively, are thought to be the young counterparts of extended radio sources. The radio emission in these objects mostly comes from lobes, which often have compact hot-spots separated by less than a kiloparsec \citep{Wilkinson1994}. For this reason, their morphology may resemble that of a gravitational lens system on scales from mas to tens of mas. 
Therefore, if not confirmed as gravitational lens candidates by up-coming observations, some of our lens candidates could be investigated as CSO candidates. In the previous section, we specified the four sources that have been previously classified as either CSO or GPS. For the remaining 36 sources no information was found.  
Another important application of this search is the possible discovery of supermassive binary black hole systems which can also display compact components on mas-scales \citep{Taylor2007,Gitti2013}.  

The confirmation of any of these sources as a millilens, would be a major discovery.
On the contrary, a null result would help to constrain $\Omega_{CO}$, as in the study of \cite{Wilkinson2001}. However, in order to constrain $\Omega_{CO}$, we need to perform such a search on a complete sample of sources. For this purpose, we selected a complete sample of sources with flux density at 8 GHz higher than 50 mJy, starting from the complete sample in CLASS \citep{Browne2003}. Sources that have not been previously observed with VLBI observations, will be observed with VLBA at 4.3 and 7.6~GHz in 2020--2021 (Project ID: BR235; PI: Readhead). This complete sample of $\sim$5000 sources, in case of null result from the search of millilenses, will help us set an upper limit on $\Omega_{CO}$ with a precision of over an order of magnitude better than in \cite{Wilkinson2001}, who analyzed a complete sample of 300 sources.

\section*{Acknowledgements}
C.C., D.B., N.M., V.P., and K.T. acknowledge support from the European Research Council (ERC) under the European Union Horizon 2020 research and innovation program under the grant agreement No 771282. V. Pavlidou acknowledges support from the Foundation of Research and Technology - Hellas Synergy Grants Program
through project MagMASim, jointly implemented by the Institute of Astrophysics
and the Institute of Applied and Computational Mathematics and by the
Hellenic Foundation for Research and Innovation (H.F.R.I.) under the "First Call
for H.F.R.I. Research Projects to support Faculty members and Researchers and
the procurement of high-cost research equipment grant" (Project 1552 CIRCE). T. H. was supported by the Academy of Finland projects 317383, 320085, and 322535.
We thank Fotini Bouzelou and John Kypriotakis who also helped in the search for lens candidates. We would like to thank the MPIfR internal referee Silke Britzen for the careful reading of the manuscript.
  We would like also to thank Leonid Petrov for maintaining the Astrogeo VLBI FITS image database and A. Bertarini, L. Vega Garcia, N. Corey, Y. Cui, L. Gurvits, X. He, Y. Y. Kovalev,
S-S. Lee, R. Lico, E. Liuzzo, A. Marscher, S. Jorstad, C. Marvin, D. Homan, M. Lister, A. Pushkarev, E. Ros, T. Savolainen, K. Sokolovski, A. Tao, G. Taylor, A. de Witt, M. Xu, B. Zhang for making VLBI images they produced publicly available. We found that a practice to upload VLBI images to a publicly available databases brings great benefits to the scientific community.
This research has made use of \textit{Aladin sky atlas} developed at CDS, Strasbourg Observatory, France \citep{Aladin2000}.

\section*{Data Availability}

The FITS images and UV-FITS files underlying this article are publicly available at the Astrogeo VLBI FITS image database (\url{http://astrogeo.org/vlbi\_images}). The usage policy of the Astrogeo VLBI FITS image database as well as the presence of unpublished material within the database does not permit the publication of either images or source related data obtained along with this study.




\bibliographystyle{mnras}
\bibliography{bibliography} 







\bsp	
\label{lastpage}
\end{document}